\documentstyle[11pt,moriond,epsfig]{article}

\def\Journal#1#2#3#4{{#1} {\bf #2}, #3 (#4)}

\def\PLB{{\em Phys. Lett.}  B}

\def\PRL{\em Phys. Rev. Lett.}

\def\PRD{{\em Phys. Rev.} D}

\def\EPJ{{\em Eur. Phys. J.} C}

\def\be{\begin{equation}}

\def\ee{\end{equation}}

\def\bea{\begin{eqnarray}}

\def\eea{\end{eqnarray}}

\begin{document}

\vspace*{4cm}

\title{$b$-HADRON PHYSICS AT LEP}

\author{ G. J. BARKER}



\address{Institute for Experimental Nuclear Physics,\\
University of Karlsruhe(TH), 76128 Karlsruhe, Germany.}

\maketitle\abstracts{A personal overview of the current status of physics
results from LEP using $b$-hadrons is presented. Emphasis is placed on
those areas where analyses are not yet finalized and there remains 
significant activity. Results are presented in the areas of $b$-quark
fragmentation, $b$-hadron lifetimes, charm counting in $b$-decays and $V_{cb}$. }

\section{$b$-Quark Fragmentation}
Modelling the fragmentation process in Monte Carlo  
is a source of systematic error to many LEP $b$-physics
analyses e.g. $\left| V_{{\mathrm ub}} \right|$. Constraining this error from
data measurements is therefore important and in the last few years precise
measurements have begun to appear e.g. from ALEPH\cite{aleph_bfrag} using the 
semi-exclusive reconstruction of $B\rightarrow D^{(*)} \ell \bar{\nu}_{\ell}$
\footnote{Unless stated otherwise, charge-conjugate states are also implied 
throughout this note.}
decays
and from SLD\cite{sld_bfrag} using a fully inclusive approach.

\begin{figure}
\begin{minipage}[t][1cm][t]{7.cm}
\includegraphics[scale=0.35]{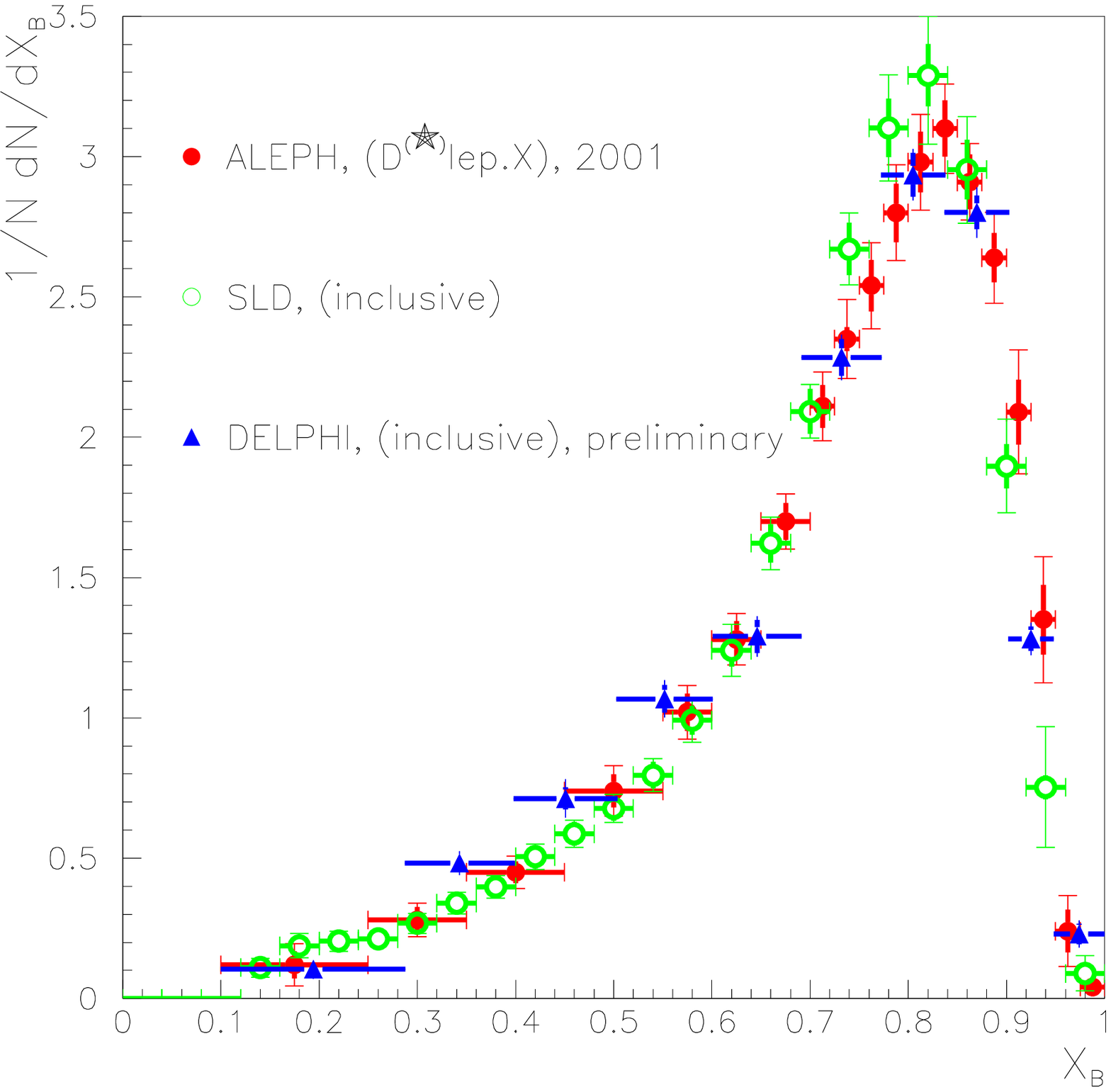}
\parbox{6.5cm}{\caption{Unfolded $f(x_B^{weak})$~distributions from ALEPH, SLD and the new
preliminary DELPHI analysis. \label{frag_world}}}
\end{minipage} 
\begin{minipage} [t][2cm][t]{8.5cm} 
\includegraphics[scale=0.6]{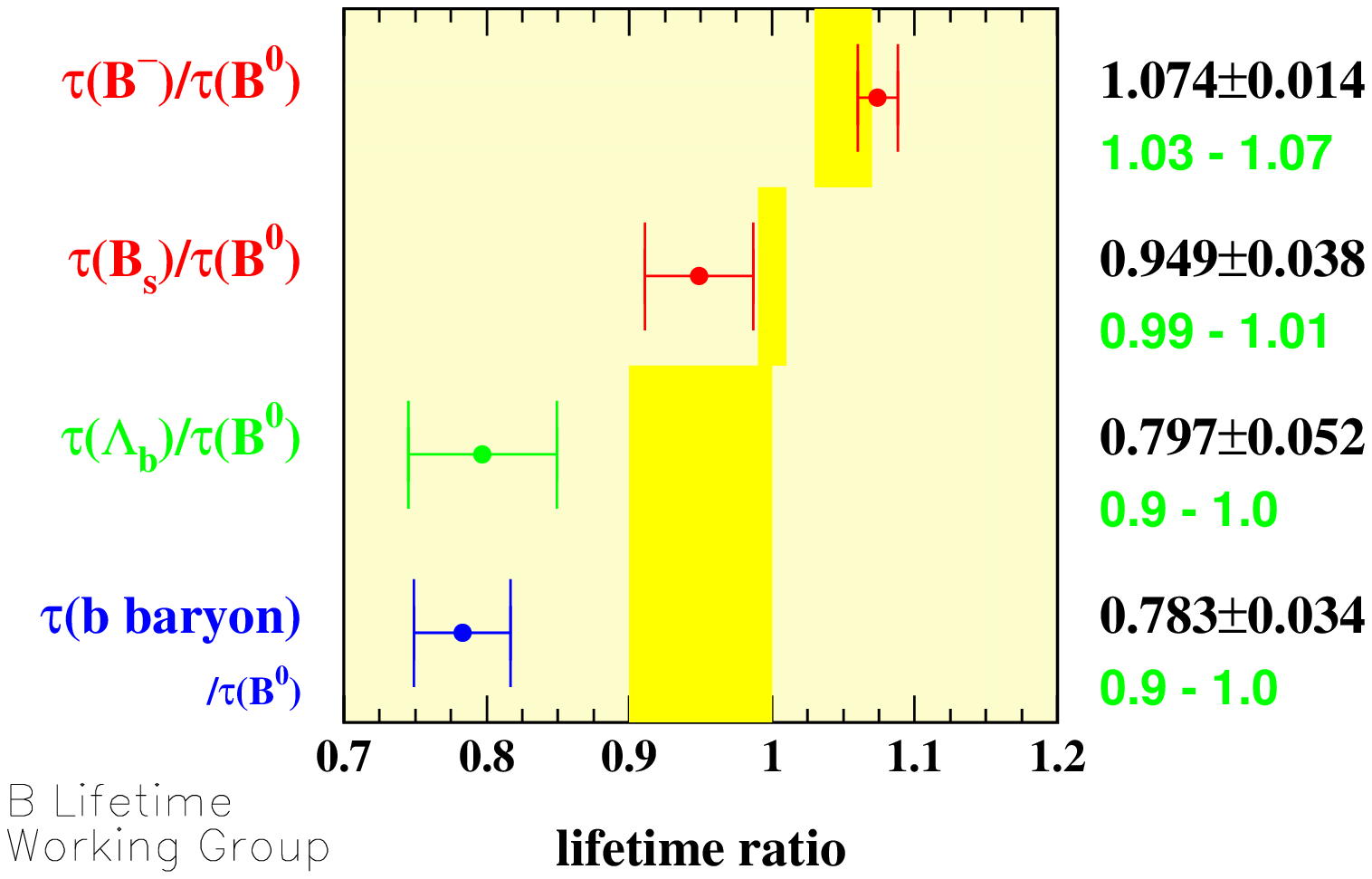}
\parbox{8.0cm}{\caption{The current world average lifetime ratios of the
different $B$-species. Results are from
the LEP experiments, SLD, CDF, Babar and Belle. `$b$-baryon' refers to 
an average over all $b$-baryon types.\label{fig:lifetimes_world}}}
\end{minipage} 
\vspace{2.cm}
\end{figure}

The fragmentation function of the $b$-quark is commonly parameterised as a
function of the variable $x_B^{weak}=E_B^{weak}/E_{beam}$~where
$E_B^{weak}$~is the energy of the weakly decaying $b$-hadron state.
Analyses
must address the difficult problem of how to unfold the underlying 
physics fragmentation function $f(x_B^{weak})$, from the measured 
$E_B^{weak}$~distribution which contains all experimental inefficiencies
and resolutions.
A new, preliminary, DELPHI analysis reconstructs inclusive $B$-decays with 
advanced neural network techniques to achieve a core $E_B^{weak}$~resolution
of $\sim 4$\% with non-Gaussian tails. The unfolding problem is solved by
employing regularisation techniques to damp down the 
oscillatory nature of the solution. The result is an extremely robust and
statistically  precise underlying  $f(x_B^{weak})$~distribution which is
presented in Figure~\ref{frag_world} along with the ALEPH and SLD result.  
\section{$b$-Hadron Lifetimes}
Recent theoretical estimates of the ratio $\tau(B^+)/\tau(B^0_d)$~predict
a lifetime difference of the order of only 2\% \cite{lenz}. This places a 
severe demand on the precision required of direct measurements if the 
framework of the Heavy Quark Effective Theory (HQET) is to be tested. 
LEP experiments have 
long dominated the $B$~lifetime world scene using mainly lepton-charm hadron
charge correlations or exclusive reconstruction to isolate the 
different b-hadron types. The most precise published
measurement of $\tau(B^0_d)$~comes from OPAL \cite{opal_b0life}
in the channel $B\rightarrow D^* \ell \bar{\nu}_{\ell}$~utilising
the charge correlation between the slow pion (from $D^*$~decay)     
and the lepton to isolate $B^0_d$. 
For $\tau(B^+)$, where the background
from other $B$-species is lower, the most precise 
measurements come from the inclusive reconstruction of
charged secondary vertices. 

In a new preliminary analysis
from DELPHI \cite{delphi_lifetime} the inclusive approach is 
taken further and neural networks are trained to identify the
$b$-hadron type based on information from the hadron fragmentation and decay
in both hemispheres of a $Z^0 \rightarrow b\bar{b}$~event.
For the case of the neural network trained to recognise
$B^+$~hadrons, a sample which is 70\%~pure with an estimated efficiency of
around 14\% is achieved.
Neural networks are also 
used to optimally estimate the $B$-momentum which, when combined with a
measurement of the distance travelled before decay, give a measure of
the $B$~proper time. 
The analysis 
reports measurements for $\tau(B^+)$~and $\tau(B^0)$~and their ratio from 
this method and the results, quoted below, are currently the most
accurate from the $Z^0$~factory experiments:  

\begin{tabbing}
ttttttttttttttttttttt\=ttttt\=ttt\=tttttt\=tttttttttttttttt\=ttttttttttttt \kill

\>  $\tau_{{\mathrm B}^+}$ \>  = \> 1.631 \> $\pm 0.012$~(stat)  \>$\pm 0.021 $~(syst)~ps \\

\> $\tau_{{\mathrm B}^0}$ \>  = \> 1.546 \> $\pm 0.018$~(stat)  \>$\pm 0.035 $~(syst)~ps \\

\>  $\frac{\tau_{{\mathrm B}^+}}{\tau_{{\mathrm B}^0}}$ \> = \>  1.054 \>$\pm 0.017$~(stat)\> $\pm 0.027$~(syst)

\end{tabbing}

The situation worldwide is summarized in
Figure~\ref{fig:lifetimes_world} which shows the current average of lifetime ratios
for the various $B$-species, 
as compiled by the $B$-Lifetime Working Group \cite{lifetimeWG}. The solid 
bands represent the theoretical expectation. It is clear that the predicted lifetime hierarchy is observed and that 
the largest discrepancy with theory comes from the $b$-baryon measurements.
It 
is rather too early to conclude whether this represents a problem 
for the current theory although results from the lattice are
beginning to appear \cite{lattice} that suggest spectator quark effects 
could account for, at most, half of the present discrepancy.

Finally, it should be noted that the $B$-factory experiments, Babar and Belle, 
currently match the precision of the best $Z^0$-decay measurements for  
$\tau(B^0_d)$~and $\tau(B^+)$~and are still statistically limited. The 
Tevatron will presumably dominate the future measurements of 
$\tau(B^0_s)$, where the most precise measurement already comes from 
CDF \cite{cdf-bslife} based on $D_s \ell$~correlations, 
and of $\tau(b-{\mathrm baryon})$~where ALEPH \cite{aleph-baryonlife}
have currently the best measurement using $\Lambda \ell$~correlations.

\section{Charm Counting}
Charm counting, i.e. the measurement of $n_c=$(the mean number of $c$~plus 
$\bar{c}$~quarks per $b$-quark decay), is important in 
attempts to resolve whether or not theoretical predictions of 
the inclusive $b$-quark semi-leptonic branching ratio ($BR_{sl}$) are
consistent with data. The most precise determinations of  $n_c$~come from 
measurements of the `wrong-sign' branching ratio 
$BR(b\rightarrow \bar{c}s(d)X)$~which, up to small corrections
for charmonium  and $b\rightarrow$no charm production, is related to
$n_c$~via $n_c \sim 1+BR(b\rightarrow \bar{c}s(d)X)$.

In a new DELPHI analysis \cite{delphi-nc}, wrong-sign $D$-mesons are 
separated from the 
`right-sign' background by utilising (a) a high performance neural network 
that tags the charge of the $b$-quark at the decay time and (b) the 
momentum spectrum of the $D$-mesons in the decaying $b$-hadron rest frame.
The preliminary results are: 
$BR(b\rightarrow \bar{D}X)=9.3 \pm 1.7(stat.) \pm 1.3 (syst.) \pm 0.4 (BR)$\%,
where $D=D^+,D^0$~and 
$BR(b\rightarrow D_s^- X)=10.1 \pm 1.0(stat.) \pm 0.6(syst.) \pm 2.8 (BR)$\%,
where in both cases the last error comes from the uncertainty on the 
branching ratio of the decay channel investigated. 

\begin{figure}[h]
\begin{minipage}[t][1cm][t]{7.5cm}
\includegraphics[scale=0.35]{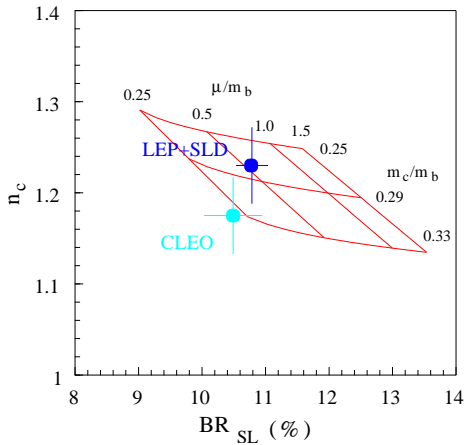}
\parbox{6.5cm}{\caption{Preliminary averages from the $Z^0$~factories (LEP+SLD) compared to
 CLEO data at the $\Upsilon(4s)$~in the $n_c$~vs $BR_{sl}$~plane.} 
\label{BRvsnc}}
\end{minipage} 
\begin{minipage} [t][2cm][t]{8.2cm} 
\vspace{-6.5cm}

Figure~\ref{BRvsnc} gives a preliminary summary of the current world measurements in the 
$n_c$~vs $BR_{sl}$~plane compared to the theoretical prediction as a function
of the $b$~and $c$~quark pole masses defined at one loop in perturbation theory.
The LEP+SLD~$BR_{sl}$~value comes from the Electroweak Working Group\cite{EWWG}
fit to the heavy flavour data and has been corrected to correspond to the 
same $b$-hadron mix as at the $\Upsilon(4s)$. The CLEO $BR_{sl}$~value comes
from~\cite{cleo_brsl}. Averages of $n_c$~were prepared by the Heavy Flavour
Working Group \cite{HFWG} and include the above DELPHI preliminary result.
One can conclude that the LEP+SLD average is compatible with the CLEO result
and that the data can be described by the theory so long as QCD corrections
are evaluated at the rather low scale of $\mu=0.35 \cdot m_b$.   
\end{minipage}
\vspace{-1.cm} 
\end{figure}
\section{$\left|V_{cb}\right|$}    
The extraction of the CKM matrix element $\left|V_{cb}\right|$~has been, and
continues to be, a very active area in LEP $b$-physics. There are two 
approaches where theoretical uncertainties are thought to be well enough under
control: The first uses the Operator Product Expansion expression linking
$\left|V_{cb}\right|$~with  $BR(b\rightarrow \ell \bar{\nu}_{\ell} X_c)$~and the 
inclusive $B$~lifetime. The second is based on the reconstruction of 
the exclusive channel $B^0_d\rightarrow D^{*-} \ell^+ \bar{\nu}_{\ell}$~and relies
on the HQET relationship for the differential cross section 
$\frac{d\Gamma}{d\omega} \propto F^2_{D^*}(\omega)\cdot
\left|V_{cb}\right|^2$. Here, $\omega=v(B^0) \cdot v(D^*)$~and
$F_{D^*}(\omega)$~is the hadronic form factor for the decay.

\begin{figure}[h]
\begin{minipage}[t][1cm][t]{6.2cm}
\includegraphics[scale=0.5]{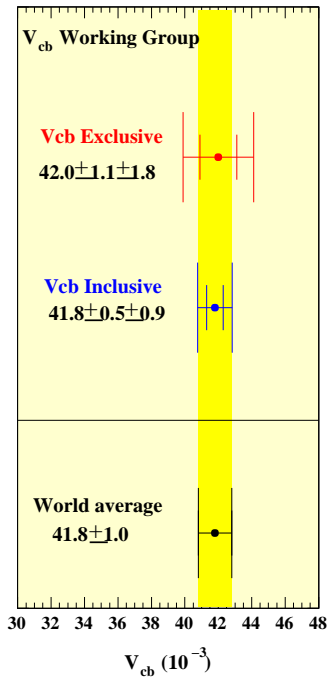}
\parbox{5.cm}{\caption{Preliminary averages for 
$\left|V_{cb}\right|$~from the exclusive and
inclusive method.}\label{vcb_world}}
\end{minipage} 
\begin{minipage} [t][2cm][t]{9.5cm} 
\vspace{-9.cm}

There are 
published results from ALEPH\cite{aleph_vcb}, and 
OPAL\cite{opal_vcb} using the exclusive approach    
where the technique is to fit the measured differential cross section 
for  $F^2_{D^*}(\omega)\cdot\left|V_{cb}\right|^2$~and then extrapolate 
the result back to the zero recoil point, $\omega=1$.  The heavy quark 
prediction for $F_{D^*}(1)$~can then be substituted to yield 
$\left|V_{cb}\right|$.

DELPHI has a new, preliminary, exclusive  measurement\cite{delphi_vcb} 
where the emphasis
has been placed on better controlling the main sources of systematic error
from their previous analysis, such as uncertainties linked to the
$\omega$~spectrum of $D^{*+}$~states coming from excited $D^{**}$~decays.
By simultaneously fitting the $D^{**}$~rate, the impact on the analysis
is greatly reduced and the result is, 
${\mathcal F}_{D^*}(1)\left|V_{cb}\right|=0.0357 \pm 0.0024(stat.) \pm
0.0018(syst.)$.

The current world status is summarised in Figure~\ref{vcb_world} 
,as compiled by the $V_{cb}$~Working Group\cite{vcbWG},
which illustrates the nice compatibility between results from the inclusive
(including LEP results only) and exclusive approach (which includes 
the preliminary DELPHI result discussed above). 
\end{minipage}
\vspace{-1.cm}
\end{figure}

\section{Conclusions}

It is fair to say that LEP has made an enormous contribution to $b$-physics
over the last ten years or so in all areas ranging from $b$-quark production
and hadronisation through lifetimes and oscillations to hadronic decay
properties and spectroscopy. Although many analyses are now finalised, 
there are still many areas of $b$-physics activity at LEP and
this review has highlighted those where new publications can be expected 
shortly.











\end{document}